\documentclass{aa}
 \usepackage{epsfig}

\begin{document}

\thesaurus{12 (12.03.4);    
           08 (08.19.4);    
           11 (11.06.1;    
              11.05.2);    
          }
\vspace{2cm}          

\title{Cosmic Star Formation and Type Ia/II Supernova Rates at high z}
\author{R. Sadat\inst{1,2}, A. Blanchard\inst{1}, B. Guiderdoni\inst{2} and J. Silk\inst{2,3} } 

\institute{ Observatoire de Strasbourg, 
            11 rue de l'Universit\'e, 67000 Strasbourg, France 
          \and 
           Institut d'Astrophysique de Paris, 98bis Bd Arago, 75014 Paris, France
          \and 
           Departments of Physics, Astronomy and Center for Particle Astrophysics Astronomy, University of California, Berkeley, USA}

\offprints{R. Sadat}

\date{Received \rule{2.0cm}{0.01cm} ; accepted \rule{2.0cm}{0.01cm} }

\maketitle

\begin{abstract}
We study how Type Ia/II supernova rates at various redshifts 
can be used to constrain the cosmic star formation rate (CSFR) history. We use a spectrophotometric model of galaxy evolution which provides a
self-consistent description of global properties of local galaxies. We have chosen two CSFR histories, with and without dust extinction, both normalized in order to reproduce the observed comoving luminosity densities at different wavelengths. We show that as expected, the SNII rates provide a direct measurement of the instantaneous CSFR, whereas the SNIa rate depends on the CSFR history during the past few Gyrs. Therefore the SNIa rate detected up to $z {\sim}$ 1 is able to provide new insight on the CSFR at higher redshift. The comparison with the data are in good agreement with both local and high-$z$ measurement and provides an additional evidence of the validity of the CSFR up to $z$ = 1, but suggests values at $z >$ 1 higher than those of Madau et al. (1996) and consistent with the presence of extinction. We also show that Type Ia supernova observations at $z{\sim}$ 1 can already put constraints on the still puzzling nature of Type Ia progenitors.   
\end{abstract}
\section{Introduction}
Supernova (SN) rates at high redshift are very important not only because of their 
ability to allow the direct determination of cosmological parameters such as 
$H_{o}$, ${\Omega}_{0}$ and ${\Omega}_{\Lambda}$, but also in understanding nucleosynthesis rates, galaxy evolution and star formation.  More specifically, the evolution of Type Ia SN rates with redshift can allow one to probe the past history of star formation in the universe, whereas their Type II SN counterparts directly probe the 
instantaneous star formation because of the short-lived massive stars which 
explode. Furthermore, observations of high-$z$ Type Ia supernovae would help us to better understand the nature of the supernova progenitors (Ruiz-Lapuente and Canal 1998). On the observational side, there has been a renewed interest in the search of SNe (Cappellaro et al. 1997 hereafter C97). An increasing number of Type Ia SNe, which are the most 
 homogeneous and most luminous SNe, are detected at redshift $z\sim$ 1 
(Perlmutter 1997, Tonry et al. 1997) and recently the Supernova Cosmology 
Project has reported the first measurement of the Type Ia SN rate at 
$z {\sim}$ 0.4 (Pain et al. 1997; hereafter P97).
Thanks to recent deep redshift surveys such as the Canada-France Survey (CFRS) and the Hubble Deep Field (HDF), it has been possible to study galaxy populations as a ``whole'' 
in order to infer the global history of star formation in the universe. More specifically, Madau et al. (1996; hereafter M96) have investigated the cosmic star and metal formation in the early universe by combining ground-based data with the HDF observations.
In this letter we use the chemical and photometric population synthesis code of 
Sadat and Guiderdoni (1998; hereafter SG98) to make predictions on the cosmic evolution of SN rates with 
redshift based on the CSFR derived from observations. We expect that such studies can provide additional constraints
on the CSFR history and metallicity evolution and will allow one to interpret SN rate measurements at high redshift which are now becoming available. 
We hereafter adopt $H_{0}$ = 50 kms$^{-1}$Mpc$^{-1}$, ${\Omega}_{0}$ = 1 and ${\Omega}_{\Lambda}$ = 0.
\section{Type Ia and II Supernovae}
The model we use is a spectrophotometric model based on stellar libraries from Kurucz (1992) supplemented by Bessell et al. (1989, 1991) for M Giants and Brett (1995) for M dwarfs.
The stars are followed along stellar tracks computed by the Geneva team (Charbonnel et al. 1996). The Initial Mass Function we choose is a ``standard'' IMF defined by:
${\phi}(m)$ $\propto$ $m^{-x}$, with index $x$=0.25 for $0.1 < m < 1 M_{\odot}$, 1.35 for ${1 < m < 2 M_{\odot}}$
and $x$ =1.7 for ${2 < m < 120 M_{\odot}}$. A more detailed description of the model is presented in a forthcoming paper (SG98). Note that, as a first step, we here restrict ourselves to models with a single metallicity (Z=Z$_{\odot}$). The chemical evolution will be studied in SG98.  
The progenitors of Type II,Ib,c SNe are stars with masses $ M > 8 M_{\odot}$. The 
rates are directly computed from the IMF. The nature of the Type Ia SNe progenitors is still a matter of debate. Indeed, it is not yet clear which of the two main competing models -- {\it double degenerate} (DD, Iben \& Tutukov 1984) or {\it single degenerate} (SD, Whelan \& Iben 1973) -- applies for the SNIa precursors. Here we assume the SD model where Type Ia SNe are produced by C-ignition and total disruption of a cold degenerate WD when this latter exceeds the Chandrasekhar mass after mass transfer in a close binary system. We estimate the rate of Type Ia SNe according to the formalism by Ferrini et al. (1992; see also Greggio \& Renzini 1983). In our SD modelling, the free parameters $m_{Binf}$ and $\alpha_{0}$, are respectively the lower limit of the total mass of the binary systems which can produce Type Ia SN, and the fraction of the total mass of 
stars which belong to these systems. They have been adjusted in order to reproduce the main properties of the solar neighbourhood. We find that the couple $[m_{Binf},{\alpha}_{0}]=[3,0.05]$ correctly reproduces the data. In order to assess the uncertainties in the nature of the progenitors, we have also used the empirical parameterization of the rates of SN Ia as derived by Ciotti et 
al. (1991; hereafter C91) $R_{Ia}\propto \theta_{SN}  t^{-s}$. $\theta_{SN}$ is the normalizing parameter. The $R_{Ia}$ evolution is controlled by the free parameter $s$. In order to account for the Fe content in clusters of galaxies, C91 concluded that $s > 1.4$. Note that our standard modelling roughly corresponds to $s \sim 1.6$. The other free parameter is the rise time which is fixed to $t_{15,0}$=0.05 in units of 15 Gyr (Renzini et al. 1993). 
\begin{figure*}[btp]
\centerline{\epsfig{file=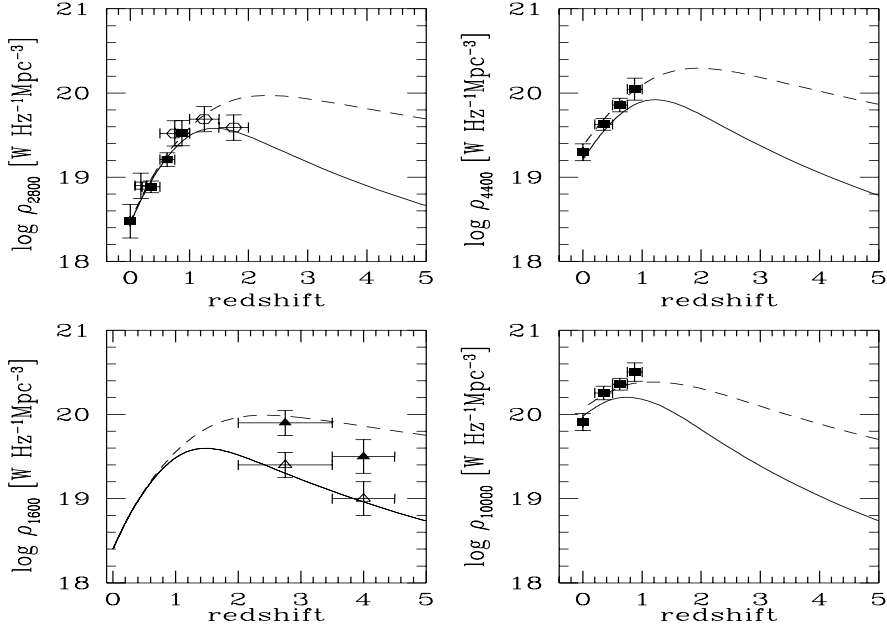,height=3.4in,width=5.3in}}
\caption{\small Evolution with redshift of the luminosity density at 
rest-frame wavelengths. The data points with error bars are taken from 
Lilly et al. (1996) ({\it filled squares}), and Connolly et al. (1997) ({\it empty 
hexagons}). {\it Open triangles} and {\it filled triangles} respectively correspond to 1600 {\AA} HDF measurements uncorrected (Madau 1997) and corrected (Pettini et al. 1997) for extinction. The {\it empty square} is from Treyer et al. (1997). The two curves correspond to the two different parameterizations M1 ({\it solid line}) and M2 ({\it dashed line}) of the CSFR.}
\end{figure*}
\section{The Cosmic Star Formation Rate}
By combining different photometric surveys such as the CFRS, up to $z$ = 1 (Lilly et al. 1995) and the HDF up to $z$ = 4 (M96), it has been possible to derive a picture of the star formation rate history of the whole 
universe (M96, Madau et al. 1997; M97). To compute the evolution of the cosmic supernova rates per comoving volume (CSNR), we make use of the CSFR as derived in M97 (model M1). We also introduce another CSFR law to account for a possible dust extinction correction (model M2). M1 and M2 have the same shape at low $z$ in agreement with the CSFR derived from observations, but at higher $z$, we have allowed the CSFR to be higher in model M2 than in M1 in agreement with Pettini et al. (1997). 
These time-dependent CSFRs are then introduced as an input in our code. The amplitude is normalized in order to match the luminosity density evolution
in the UV--continuum with the following constants of proportionality between the observed L$_{UV}$ and our derived SFR: ($7.5\times 10^{19}, 6.5\times 10^{19}$)
 in units of W Hz$^{-1}$ (M$_{\odot}$ yr$^{-1}$)$^{-1}$ at (1500 \AA, 2800 \AA). Before computing the SN rates, we first have checked whether our simple 
model with a ``standard'' IMF and the parameterization of the CSFR is able to reproduce the observed volume averaged density luminosities $\rho_{\nu}$ at different ${\lambda}$. Figure 1 shows that the agreement at longer ${\lambda}$ is satisfying. The 4400 and 10000 {\AA} luminosity densities seem to hint at model M2 in agreement with the IR/submm background which also suggests the presence of extinction at high $z$ (Guiderdoni et al. 1997).  
\section {Cosmic Evolution of SN Rates}
\begin{figure}[btp]
\centerline{\epsfig{file=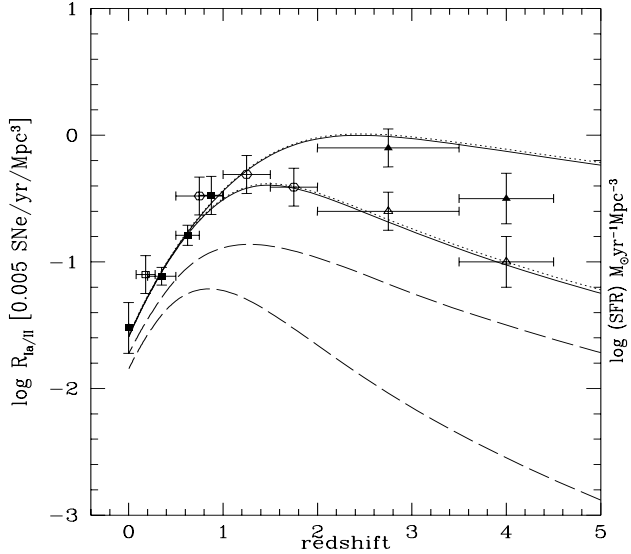,height=3.0in,width=3.5in}}
\caption{\small Evolution of SN rest-frame rates per comoving volume for model M1 ({\it lower curves}) and M2 ({\it upper curves}): $R_{II/Ib,c}$ ({\it solid lines}); $R_{Ia}$ ({\it dashed lines}). CSFRs are shown as {\it dotted lines}. Data are derived from the UV fluxes with the constants of proportionality mentionned in section 3.}
\end{figure}
\noindent
In this section, we use the CSFR to derive the CSNR for both II/Ib,c and Ia Types. To our knowledge, this is the first time that the evolution of SN rates with redshift is predicted from a self-consistent spectro-photometric modelling of galaxy
evolution at high $z$ and independently of the details of individual galaxy evolution. As already mentionned, the direct measurement of SNe can be used as an independent test for the cosmic star and metal formation in the universe. In figure 2, we have plotted the predicted evolution of the CSNR per unit of comoving volume 
with redshift. The Type II/Ib,c rate shows the same shape as the 
instantaneous CSFR, that is the rise, peak and drop from high redshift to the present time. This means that the SNII/Ib,c rate can be used as an independent tracer of the star formation rate. The Type Ia rate $R_{Ia}$ has a different shape from the CSFR while coincidently it has nearly the same behaviour as the B-band luminosity. The most important point is the time delay we observe between the stellar (binary in this case) birth and the explosion time. The occurrence of the SNIa rate peak is shifted by a few Gyrs relatively to the CSFR peak. However, at $z \ge$ 0.9, Type Ia SN can be used as a probe of the past history of the CSFR. Furthermore, as can be seen from figure 3, $R_{Ia}$ has very distinct shape (different normalization and different time of peak occurrence) depending on the adopted CSFR shape, and this difference is higher at higher $z$. Therefore, measurements of SNIa at $z {\sim}$ 1 would be able to discriminate between the models. 

\section{Comparison with observations}
\begin{figure}[btp]
\centerline{\epsfig{file=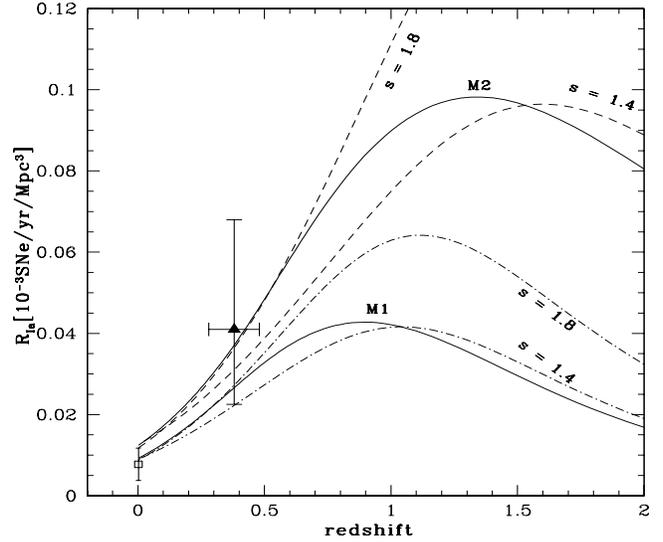,height=3.0in,width=3.5in}}
\caption{\small Predicted evolution of the rest-frame Type Ia SN rate per comoving volume. The curves labelled with ${s}$ correspond to $R_{Ia}$ as derived assuming the C91 parameterization normalized in order to reproduce the present day value for the same CSFR. The horizontal bar is the observational redshift range.}
\end{figure}
\begin{figure}[btp]
\centerline{\epsfig{file=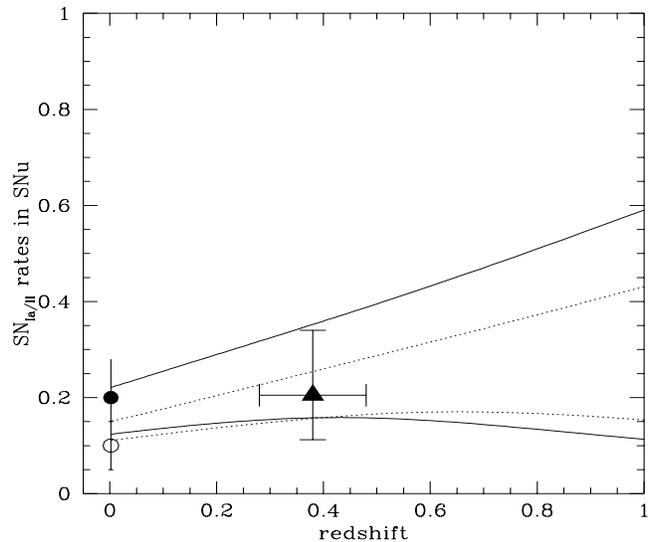,height=3.0in,width=3.5in}}
\caption{\small Evolution of SNII ({\it upper curves}) and 
SNI{a} ({\it lower curves}) rates for models M1({\it solid lines}) and M2 ({\it dotted lines}) compared to the observations. The {\it filled circle} and {\it open circle} respectively correspond to SNII/Ib,c and SNIa rates from C97, and the {\it filled triangle} is from P97.}
\end{figure}
\noindent
Supernova rate has been measured in nearby galaxies 
by several authors (Cappellaro \& Turatto 1988, Evans et al. 1989, C97). At higher redshift,
Pain and his collaborators have recently reported the first high-$z$ ($z \sim$ 0.4) 
rest-frame Type Ia SN rate with a value of ${\sim}$ 0.82 h$^{2}$ SNu. In order to compare the model to the observations, we have converted the observed rate in SNu (SNe/100 yr/ 10$^{10}$ L$_{B\odot}$) into a rate in yr$^{-1}$ using blue luminosities as computed by our code. From figure 3, we can see that although the available measurement of the SNIa rate does not allow one to discriminate between models M1 and M2, the redshift is still low and the statitics is poor. Measurement at $z \sim$ 1 with the same statistics, would begin to discriminate between models, independently of the adopted model for Type Ia progenitors. By investigating different models for SNIa rates using C91 empirical parameterization (see figure 3 caption), we have found that a comparison between these model predictions and observations at low and intermediate $z$ shows that the best match is obtained for models with a rapidly declining Type Ia rate ($s > 1.4$) consistent with the constraints on $s$ from chemical evolution. In figure 4 we have plotted the predicted SNIa/II rates in SNu. As can be seen, the SNIa rate is nearly constant. This is expected since both the $R_{SNIa}$ per unit of comoving volume and the B-band luminosity have the same shape. Therefore the peak observed in $R_{SNIa}$ has 
been erased, leading to a constant ratio. However, this is 
not true for the Type II SN rate which turns out to be a monotonic function increasing with redshift. This is a consequence of the fact that the stellar population which emits the bulk of the B-band luminosity is older than the SNII progenitors. Note that the SN rates in 
SNu do note change drastically with the adopted CSFR prescription. This means that SN rates in SNu loose a large part of the information on the absolute value of the CSFR. Our model reproduces fairly well the observed local Type Ia rate and our prediction for high-$z$ rates is in good agreement (within the error bars) with the observed Type Ia rate of P97. At $z \sim$ 1 we expect that the SNIa rate in SNu does not change from its local value while the SNII rate is expected to be 3 times the local value. This leads to the decrease of the Type Ia/II SN ratio from $z$=0 to $z$=1 (see also Yungelson \& Livio 1997). However, the Type II SN rate is not observed yet and this prediction could be confirmed with future instruments such as the NGST. 
\section{Conclusion}
We have computed the CSNR from the CSFR using a self-consistent modelling. We found that:
\begin{itemize}
\item{ The adopted standard IMF allows one to fit the observed local and high-$z$ colours of the universe. }
\item{Type II SN rates are very good tracers of instantaneous star formation and could be used to set constraints on the CSFR at $2 < z < 4$ with the new generation of telescopes such as the NGST.}
\item {Type II SN rates $R_{SNII}$ (expressed in SNu) are an increasing function with redshift while the $R_{SNIa}$/$R_{SNII}$ ratio is decreasing.}
\item{Our model predictions for SN rates are in a good agreement with observations locally as well as at higher $z$, and consistent with the current limits on the CSFR.} 
\item{At redshift $z \sim 1$, a comparison between the predicted and the observed Type Ia rates
would allow us to put stong constraints on models for SN Ia progenitors. In the frame of C91 models, we found that the best match is obtained for models with a steep evolution of SNIa rate (i.e. $s > 1.4$) consistent with the chemical evolution of Fe in clusters of galaxies.}  
\item{Intermediate and high-$z$ Type Ia SN rates can be used as an independent test to constrain the CSFR. The current data at $z \sim 0.4$ are not sufficient to disentangle between the models but observations of Type Ia supernovae at $z\sim 1$ would be critical for understanding the CSFR and would allow one to assess whether the $z > 1$ CSFR is higher than directly observed from the UV.}

\end{itemize}

\begin{acknowledgements}
We thank R. Pain and R. Mochkovitch for fruitfull comments on supernova observations and theory. JS and RS acknowledge support from ``The chaire Blaise Pascal''.
\end{acknowledgements}
\end{document}